\documentclass[epj]{svjour}

\usepackage{amssymb}
\usepackage{amsmath}
\usepackage{graphicx}
\usepackage{bm}
\usepackage{widetext}

\begin{document}

\title{Momentum of charge-magnetic coil systems}

\author{Francis Redfern}
\institute{Texarkana College, Texarkana, Texas 75599}

\abstract{
Solenoids and toroidal solenoids exposed to an electric field have been thought
to contain hidden momentum. Here I show that there is no hidden momentum in
these systems.}


\maketitle

\section{Introduction}

In 1891 J.J. Thompson pointed out an apparent paradox \cite{JJ91} where
electromagnetic systems at rest could contain non-zero electromagnetic
momentum. This result languished for many years until, in 1967, Shockley
and James \cite{Shock} examined a charge-magnet system and claimed it
had to contain a form of mechanical momentum they termed ``hidden
momentum'' in order for momentum conservation not to be violated. A year
later Coleman and Van Vleck \cite{Coleman} published a detailed analysis
on the question of a point charge in the vicinity of a magnet and
concluded that the conjecture of Shockley and James was correct.

Since these papers were published, charge-magnet systems have been thought
to contain this new form of linear mechanical momentum. (Also, there is supposed
to be hidden angular momentum in many of these systems.) I have published work
on several charge-magnet systems that shows they contain no such momentum
\cite{redfern1,redfern2}. What has been ignored are the electromagnetic forces
that arise when an electric field is applied to a magnet or a magnet is formed
in an electric field. When these systems originate, mechanical momentum due to
these forces, as well as electromagnetic momentum, is imparted to them such that
you can't consider them to be at rest in the reference frame in which they
originated. Ignoring the appearance of mechanical momentum has resulted in the
invention of hidden momentum, which I argue is an unphysical concept. When
forces of origination are taken into account, it is clear that the momentum
that is stored in the electromagnetic field is equal and opposite to the
mechanical momentum the system gains as a result of those forces. If the system
(or parts thereof) is kept stationary, the so-called hidden momentum actually
resides in the environment responsible for keeping the system at rest.

In the following sections I will show that systems consisting of a charge in the
vicinity of a solenoid and toroidal solenoid contain no hidden momentum.
I examine several
cases. First I look at the situation where a charge is in place near the coil
(in the geometrical center of the coil in the case of the toroid) and magnetic
flux is built up in the coil. Then I look at cases where a charge is moved
from infinity to the vicinity of coils already containing flux. In none
of these cases is hidden momentum necessary for momentum conservation.

\section{Activating a solenoid near a charged particle}

Consider a charge $q$ a distance $R$ from a long air solenoid. Both charge and
coil are at rest and current is flowing in the solenoid. Assuming no electrical
shielding, there is an electric
field inside the solenoid due to the presence of the charge, thus there should
be an electromagnetic momentum given (in the Lorentz formulation) by the formula
\begin{equation}
\label{em momentum}
P_{em} = \epsilon_o\int_V(\bm{E}\times\bm{B})dV = \frac{q}{4\pi}\int_V\frac{\bm{\hat{r}}}{r^2}\times B\bm{\hat{k}}dV,
\end{equation}
where $\bm{r}$ is the position vector from the charge to a point on the axis of
the solenoid. Have the solenoid be centered on the $z$ axis with its
magnetic field in the positive $z$ direction and the charged particle along the
$x$ axis at $x = -R$. In this configuration $\bm{\hat{r}}\times\bm{\hat{k}} =
-cos\alpha\bm{\hat{j}}$, where the angle $\alpha$ is measured from the $x$ axis
to $\bm{r}$, $a$ is the radius of the solenoid ($a << R$), and $dV =
\pi a^2 dz$. With these expressions, the electromagnetic momentum becomes
\begin{equation}
\label{system em momentum}
P_{em} = -\frac{q\pi\Phi_B}{4\pi R}\bm{\hat{j}}\int_{-\pi/2}^{\pi/2}cos\alpha d\alpha = -\frac{q\Phi_B}{2\pi R}\bm{\hat{j}},
\end{equation}
where $\Phi_B = \pi a^2 B$ is the magnetic flux inside the solenoid.

Since the system is at rest, there appears to be no mechanical momentum present
to balance the electromagnetic momentum. To conserve momentum the conventional
view is that there is hidden momentum in the solenoid, equal and opposite to
the electromagnetic momentum. However, consider the following scenario.

Have the charge positioned as before, but let there be no current in the
solenoid. Let a current build up in the solenoid such that the flux in the coil
goes from zero to $\Phi_B$. Faraday's law says there will be an induced electric
field at the position of the charge given by
\begin{equation}
\label{faraday}
\bm{E} = -\frac{\partial\bm{A}}{\partial t},
\end{equation}
where $\bm{A}$ is the vector potential at that point. The vector potential
for a solenoid can be found as follows.
\begin{equation}
\label{B flux}
\Phi_B = \oint\bm{B}\cdot d\bm{S} = \oint\nabla\times\bm{A}\cdot d\bm{S} = \oint\bm{A}\cdot d\bm{l} = 2\pi rA.
\end{equation}
In the above equation the first integral is a surface integral over the
cross-sectional area of the solenoid, the second uses the fact that $\bm{B} =
\nabla\times\bm{A}$, and the third, a line integral, follows from Stokes'
theorem. The final result follows from symmetry: The right hand rule shows that
if the magnetic field of the solenoid is in the positive $z$ direction, the
vector potential is in the $\bm{\hat{\phi}}$ direction, where $\bm{\hat{\phi}} =
-sin\phi\bm{\hat{i}} + cos\phi\bm{\hat{j}}$, so that
\begin{equation}
\label{A for solenoid}
\bm{A} = \frac{\Phi_B}{2\pi r}\bm{\hat{\phi}} = -\frac{\Phi_B}{2\pi r}\bm{\hat{j}},
\end{equation}
where $\bm{\hat{\phi}} = -\bm{\hat{j}}$ at the position $(-R,0,0)$ of the
charge.

The electric field acting on the charge a distance R from the coil is
\begin{equation}
\label{E on q}
\bm{E} = \frac{\dot{\Phi_B}}{2\pi R}\bm{\hat{j}},
\end{equation}
with $\dot{\Phi_B} = \partial\Phi_B/\partial t$.
With an electric field acting on the charge as the magnetic flux increases,
there must be an equal and opposite force if the charge is to be held
stationary. This force is
\begin{equation}
\label{opposing force}
\bm{F} = -\frac{q\dot{\Phi_B}}{2\pi R}\bm{\hat{j}}.
\end{equation}
As a result the external agent necessary to supply this force will gain an
amount of mechanical momentum given by
\begin{equation}
\label{Pme solenoid}
\bm{P}_{mm} = \int\frac{q\dot{\Phi_B}}{2\pi R}\bm{\hat{j}}dt =\frac{q\Phi_B}{2\pi R}\bm{\hat{j}}.
\end{equation}
Note that this is equal and opposite to the electromagnetic momentum contained
in the solenoid. Hence the total momentum of the system -- charge, solenoid, and
external agent -- is conserved without the need for a hidden form.

\section{Activating a toroidal solenoid near a charged particle}

Consider a current loop of radius $a$ lying in the $x$-$y$ plane and
centered at the origin with the $z$ axis marking the axis of symmetry of the
loop. According to Jackson \cite{Jackson5}, the vector potential a distance $r$
from the center of the loop in the $x$-$y$ plane ($\theta = \pi/2$) and where
$a<<r$ is approximately, in SI units,
\begin{equation}
\label{A for loop}
\bm{A} = \frac{\mu_o Ia^2}{4r^2}\bm{\hat{\phi}}.
\end{equation}
The vector potential at the location $(R, 0, 0)$ is given by the above equation
with $\bm{\hat{\phi}} = \bm{\hat{j}}$. Imagine this loop is an element of a
toroidal solenoid with $N$ turns whose center is a distance $R$ from the loop.
The coil will be oriented such that its symmetry axis will be in the $z$
direction and with a current such that the magnetic field is in the
$\bm{\hat{\phi}}$ direction. The vector potential due to all the current loops
in the coil will now be directed in the positive $z$ direction. Since each loop
contributes the same quantity to the vector potential along the coil's symmetry
axis, the vector potential along the axis will be
\begin {equation}
\label{Az}
A_z = \frac{\mu_o NIa^2}{4r^2}.
\end{equation}
Along this axis $r = R/cos\theta$ such that $A_z$ becomes
\begin{equation}
\label{Az 2}
A_z = \frac{\mu_o NIa^2}{4R^2}cos^2\theta.
\end{equation}
At the position of the charge, $\theta = 0$ so that
\begin{equation}
\label{Az 3}
A_z = \frac{\mu_o NIa^2}{4R^2} = \frac{\Phi_B}{2R},
\end{equation}
since $B = \mu_o NI/2\pi R$ inside the coil and the cross-sectional area of the
coil is $\pi a^2$, again taking the radius of the coil $a << R$. Following the
same procedure as for the solenoid, it is easily seen that the electromagnetic
linear momentum in the coil is
\begin{equation}
\label{Pem toroid}
\bm{P}_{em} = \frac{q\Phi_B}{2R}\bm{\hat{k}},
\end{equation}
whereas the mechanical momentum obtained by an external agent required to keep
the charge at rest while the flux in the coil increases is
\begin{equation}
\label{Pme toroid}
\bm{P}_{mm} = -\frac{q\Phi_B}{2R}\bm{\hat{k}}.
\end{equation}
Once again, the total momentum -- the mechanical momentum of the external agent
plus the electromagnetic momentum -- is zero such that no hidden momentum is
necessary to achieve momentum conservation.

\section{Bringing a charged particle into the vicinity of a solenoid}

Placing a charged particle near a coil and turning on the electricity is
not the only way to put together a charge-magnetic coil system and derive the
momentum changes involved. You can also move the charge from a sufficiently
great distance, where there is no interaction between the charge and coil, to
the vicinity of the coil.

I will now show that there is no electric or magnetic field to act on
the charged particle as it moves in a region external to an infinite solenoid
or a toroidal solenoid. The potential four-vector is given in general by
\begin{equation}
\label{4 vector}
A^{\mu'} = (A_{x'}, A_{y'}, A_{z'}, V'/c),
\end{equation}
where $\bm{A}'$ is the three-vector potential and $V'$ is the scalar potential.
Since there is no net charge on the solenoid, the scalar potential is zero.

It can be shown that, in a certain reference frame where the scalar
potential is zero (or constant), the vector potential is not a function of time
(no electric field), and the curl of the vector potential is zero (no magnetic
field), there can be no electric or magnetic field in any other inertial
reference frame. This follows from the fact, taking $\bm{v}$ arbitrarily to be
in the positive $x$ direction, that $\bm{A}$ is a function of $\gamma(x-vt)$.

Let $\bm{A}'$ be a position-dependent but time-independent vector potential in a
certain (primed) frame of reference; that is, a function of $(x',y',z')$ but
not $t'$; and let $V' = 0$. Transform the primed frame to the lab (unprimed)
frame by having it move with speed $v$ in the positive $x$ direction. In the
lab frame the four-vector potential will be
$(A_x,A_y,A_z,V/c) = (\gamma A_{x'},A_{y'},A_{z'},\gamma(v/c)A_{x'})$,
where $x' = \gamma (x-vt)$ and $V$ is the scalar potential. The electric field
in the lab frame will be given by
\begin{equation}
\label{lab E field}
\bm{E} = -\frac{\partial A_x}{\partial t}\bm{\hat{i}} -
\frac{\partial A_y}{\partial t}\bm{\hat{j}} -
\frac{\partial A_z}{\partial t}\bm{\hat{k}} - \nabla V.
\end{equation}
Note that
\begin{eqnarray}
\label{dA/dt}
\frac{\partial A_x}{\partial t} &=& \gamma\frac{\partial A_{x'}}{\partial x'}\frac{\partial x'}{\partial t}
= -\gamma^2v\frac{\partial A_{x'}}{\partial x'}, \nonumber \\
\frac{\partial A_y}{\partial t} &=& \frac{\partial A_{y'}}{\partial x'}\frac{\partial x'}{\partial t}
= -\gamma v\frac{\partial A_{y'}}{\partial x'}, \nonumber \\
\frac{\partial A_z}{\partial t} &=& \frac{\partial A_{z'}}{\partial x'}\frac{\partial x'}{\partial t}
= -\gamma v\frac{\partial A_{z'}}{\partial x'}.
\end{eqnarray}
Also
\begin{eqnarray}
\label{grad V}
\frac{\partial V}{\partial x} &=& \gamma v\frac{\partial A_{x'}}{\partial x'}\frac{\partial x'}{\partial x} = \gamma^2 v\frac{\partial A_{x'}}{\partial x'}, \nonumber \\
\frac{\partial V}{\partial y} &=& \gamma v\frac{\partial A_{x'}}{\partial y'}, \nonumber \\
\frac{\partial V}{\partial z} &=& \gamma v\frac{\partial A_{x'}}{\partial z'},
\end{eqnarray}
where $y$ and $z$ have been replaced by $y'$ and $z'$ for clarity in recognizing
that substituting the quantities in the above two sets of equations into the
electric
field of Eq. (\ref{lab E field}) and using the fact that $\nabla'\times\bm{A}'
= 0$ results in the electric field in the lab frame (and in any other inertial
frame, since the lab frame is arbitrary) being zero. The magnetic field is also
zero since
$\bm{B} = \bm{v}\times\bm{E}/c^2$. A moving charge will therefore experience no
electromagnetic force in either the vicinity of a long solenoid or a toroidal
solenoid (with no magnetic field leakage).

Next, look at the effect the moving charge has on the solenoid. In
\cite{redfern4} I showed that a charge moving toward a current loop in plane of
the loop will impart an impulse to the loop given by
\begin{equation}
\label{impulse to current loop}
\Delta\bm{P} = \frac{\mu_o q IA}{4\pi r^2}\bm{\hat{j}} = -\frac{1}{2}\mu_o\epsilon_o\bm{E}\times\bm{m},
\end{equation}
where $\bm{m}$ is the magnetic moment of the current loop.
The loops of a solenoid will be off-center from the direction of motion of the
charge, so a new calculation taking this into account is in order. The
electric field of moving charge, $q$, at point $\bm{r}$ in the lab frame is
given by the following \cite{RindlerE}.
\begin{equation}
\label{rel E field}
\bm{E} = \frac{1}{4\pi\epsilon_o}\frac{q\bm{r}}{\gamma^2 r^3(1-\beta^2sin^2\alpha)^{3/2}},
\end{equation}
where $\beta = v/c$ and $\alpha$ is the angle between the direction of motion of
the charge and $\bm{r}$. The motion of the charge will produce a magnetic field
given by
\begin{equation}
\label{rel B field}
\bm{B} = \frac{1}{c^2}\bm{v}\times\bm{E} = \frac{1}{4\pi\epsilon_o}\frac{q\bm{v}\times\bm{r}}{\gamma^2c^2r^3(1-\beta^2sin^2\alpha)^{3/2}},
\end{equation}
where $\bm{v}$ is the velocity of the charge. If the charge is moving in the
positive $x$ direction, $\bm{v}\times\bm{r} = v(-z\bm{\hat{j}} +
y\bm{\hat{k}})$.

The solenoid is stationary in the lab frame with its axis parallel to the $z$
direction and its magnetic field in the positive $z$ direction. Consider a
current loop of the coil with its center located at the origin of a cylindrical
coordinate system. The current density of the loop will be
\begin{equation}
\label{J of loop}
\bm{J} = J\bm{\hat{\phi}} = I\delta(\rho-a)\delta(z)(-sin\phi\bm{\hat{i}} + cos\phi\bm{\hat{j}}).
\end{equation}
Here, the Dirac delta function is used, $\rho$ is the cylindrical radial
coordinate, $a$ is the radius of the loop, and $\phi$ is the azimuthal
coordinate. The center of the current loop is located at the position $\bm{r} =
(R,0,h)$ in a Cartesian coordinate system centered on the charge. An
element of current on the loop is located at $(R+acos\phi,asin\phi,h)$. In the
slow motion situation, $\gamma$ can be taken as one and $\beta \approx 0$. (This
will be the case for the rest of this article.) The force per unit volume
acting on the loop is then approximately given by
\begin{eqnarray}
\label{F/V on loop}
\bm{f} &=& \bm{J}\times\bm{B} \\
&=& \frac{1}{4\pi\epsilon_o}\frac{qvI\delta(\rho-a)\delta(z)}{c^2r^3}(asin\phi\bm{\hat{i}} + asin^2\phi\bm{\hat{j}} + hsin\phi\bm{\hat{k}}. \nonumber
\end{eqnarray}
It is tempting to take $r \approx \sqrt{l^2 + h^2}$ when $a << r$, but this
makes the magnetic field larger at locations on the loop farther away from
the charge. The proper way to do the calculation is to maintain the accuracy of
 $r^3$. However, you can expand $r^{-3}$ and drop terms the order of $a^2/r^2$.
It turns out this is not necessary here, since either side of the loop is the
same distance from $q$. This will not be the case when the toroidal solenoid is
considered later on.

The magnetic field forms circular loops centered on the line
of action of the velocity of the charge. If you use the thumb of your right hand
to point in the direction of the charge's motion, then your curled fingers will
represent the magnetic field lines. Note that there will be a torque acting on
a current loop on the positive $z$ axis with the torque vector in the same
direction as
the velocity of the charge. For a current loop on the negative $z$ axis the
torque is in the opposite direction such that no net torque is applied to an
infinite solenoid or to a charge approaching the halfway point of a finite
solenoid, although the coil will experience stress as the charge moves. An
off-center approach of the charge toward a finite solenoid will produce a net
torque on the coil.

Realizing that you don't need to engage in the extra effort required to expand
$r^{-3}$, the force acting on the loop is
\begin{equation}
\label{F on loop}
\bm{F} = \int_V\bm{f}\rho d\rho dz d\phi = \frac{1}{4\pi\epsilon_o}\frac{qvI\pi a^2}{c^2 r^3}\bm{\hat{j}}.
\end{equation}
To get the impulse applied to the current loop, you must move the charge
in from infinity in a direction perpendicular to the solenoid axis and directly
at it. Take $R$ as the final distance from the charge to the solenoid. Replace
$h$ with $z$, the location of a loop from the intersection of the line of action
of $\bm{v} = v\bm{\hat{i}}$ and the axis of the solenoid as this will later
become a variable. The speed can be given in terms of the time rate of change of
the angle $\alpha$, the angle between $\bm{v}$ and $\bm{r}$, as $v =
z/(cos^2\alpha)(d\alpha/dt)$. Also note that $r = z/sin\alpha$. The force on the
current loop is then
\begin{equation}
\label{F on loop 2}
\bm{F} = \frac{1}{4\pi\epsilon_o}\frac{qI\pi a^2}{c^2 z^2}sin\alpha\frac{d\alpha}{dt}\bm{\hat{j}}.
\end{equation}
The impulse delivered to the current loop will follow from the integral of
$\bm{F}dt$ over the distance traveled by the charge. Let $\alpha_R$ be the value
of $\alpha$ when the charge reaches the distance $R$ from the axis of the coil.
The mechanical momentum received by the loop is then
\begin{eqnarray}
\label{P_loop}
\bm{P}_{loop} &=& \frac{1}{4\pi\epsilon_o}\frac{qI\pi a^2\bm{\hat{j}}}{c^2 z^2}\int_0^{\alpha_R}sin\alpha d\alpha \nonumber \\
&=& \frac{1}{4\pi\epsilon_o}\frac{qI\pi a^2\bm{\hat{j}}}{c^2 z^2}(1 - cos\alpha_R).
\end{eqnarray}

Next, you turn this equation into another differential equation for the momentum
received by the entire solenoid. To do this you can replace $\alpha_R$ with
$\alpha$, now a variable, and $I$ by $dI = Kdz$ where $K = NI/L$ is
the current per unit length of a solenoid of length $L$ with $N$ turns. Also,
to perform the integral involved, you make the substitutions $z = Rtan\alpha$
and $dz = Rd\alpha/(cos^2\alpha)$. The loop now contributes an amount of
mechanical momentum $d\bm{P}_{mm}$ to the coil of
\begin{equation}
\label{dP_mm}
d\bm{P}_{mm} = \frac{1}{4\pi\epsilon_o}\frac{qK\pi a^2\bm{\hat{j}}}{c^2 R tan^2\alpha cos^2\alpha}(1-cos\alpha)d\alpha.
\end{equation}
To integrate this note that $cos^2\alpha tan^2\alpha = sin^2\alpha$ and that
$(1-cos\alpha)d\alpha/sin^2\alpha = d\alpha/(1+cos\alpha) =
d\theta/cos^2\theta = d(tan\theta)$, where $\theta = \alpha/2$. The integral is
taken from $\theta = -\pi/4$ to $\theta = \pi/4$. The result is
\begin{equation}
\label{P_mm}
\bm{P}_{mm} = \frac{q\Phi_B}{2\pi R}\bm{\hat{j}},
\end{equation}
where $\Phi_B = \mu_o K\pi a^2 = \mu_o NI\pi a^2/L = \pi a^2 B$, and where
$c^{-2} = \epsilon_o\mu_o$ has been used. Note that this mechanical momentum
applied to the solenoid is equal and opposite to the electromagnetic momentum
in the coil given by Eq. (\ref{system em momentum}). Again, no hidden momentum
is necessary to preserve momentum conservation as the total momentum is zero.
\hfill\break

\section{Bringing a charged particle into the vicinity of a toroidal solenoid}

For this calculation the charge is moved from infinity to a point in the plane
of and outside a current loop, and then $N$ loops are arranged to form the
toroid. You saw in the previous section that the
electric field acting on the charge has to be zero, so the only mechanical
momentum is that imparted to the loop. The magnetic field and current density
are given by Eqs. (\ref{rel B field}) and (\ref{J of loop}) as before, but now
the loop is at the position $\bm{r} = (s,R,0)$ with respect to the charge, where
$s$ is originally very far away and goes to zero where the charge finally comes
to rest. $R$ will turn out to be the radius of the toroid.

In this instance you cannot take $r$ to be the distance from the charge to the
center of the current loop when calculating the force on the loop. Instead,
$\bm{r} = (s+acos\phi, R+asin\phi, 0)$ must be used in the denominator, but you
can expand the denominator since it is assumed that $a << R$. Following standard
expansion procedure, you can write $r^{-3}$ as
\begin{equation}
\label{1/r^3}
\frac{1}{r^3} = \frac{1}{r_o^3}\left[1 - \frac{3a(s cos\phi + R sin\phi)}{r_o^2}\right],
\end{equation}
where $r_o = \sqrt{s^2 + R^2}$. The magnetic field due to the motion of the
charge is then
\begin{eqnarray}
\label{expanded B}
\bm{B} &=& \frac{1}{4\pi\epsilon_o}\frac{qv(R+asin\phi)\bm{\hat{k}}}{c^2 r_o^3}
\nonumber \\
&\times& \left[1 - \frac{3a(s cos\phi + R sin\phi)}{r_o^2}\right].
\end{eqnarray}
The force per unit volume on the loop is given by
\hfill\break
\hfill\break
\hfill\break
\hfill\break
\begin{widetext}
\begin{equation}
\label{F/V toroidal loop}
\bm{f} = \bm{J}\times\bm{B} = \frac{1}{4\pi\epsilon_o}\frac{qvI\delta(\rho-a)\delta(z)(R+asin\phi)}{c^2 r_o^3}\left[1 - \frac{3a(scos\phi+Rsin\phi)}{r_o^2}\right](cos\phi\bm{\hat{i}}+sin\phi\bm{\hat{j}}).
\end{equation}
\end{widetext}
Performing the volume integral of the above equation yields the force on the
current loop. Only those terms that aren't zero when the integral over $\phi$ is
taken are shown in the equation below.
\begin{eqnarray}
\label{F on toroid loop}
\bm{F} &=& \frac{1}{4\pi\epsilon_o}\frac{qvNI\pi a}{c^2 r^3} \\
&\times& \int_0^{2\pi}\left[asin^2\phi\bm{\hat{j}} - \frac{3a}{r_o^2}Rscos^2\phi\bm{\hat{i}} - \frac{3a}{r_o^2}R^2sin^2\phi\bm{\hat{j}}\right]d\phi. \nonumber
\end{eqnarray}
When $N$ current loops are put together to form the toroid, the contributions
to the force acting on the coil in the $y$ direction cancel, leaving only the
contribution in the $x$ direction. Now, I permute the coordinate axes to bring
the axis of symmetry of the coil to coincide with the $z$ axis. The magnetic
field is then in the $\bm{\hat{\phi}}$ direction. The result of the integration
is
\begin{equation}
\label{F on toroid}
\bm{F} = -\frac{1}{4\pi\epsilon_o}\frac{3qviNI\pi a^2Rs}{c^2r_o^5}\bm{k}.
\end{equation}
The speed of the charge in terms of $s$ is $v = -ds/dt$ such that the impulse
delivered to the coil is given by
\begin{equation}
\label{P_mm t-coil integral}
\bm{P}_{mm} = -\frac{1}{4\pi\epsilon_o}\frac{qNI\pi a^2R\bm{\hat{k}}}{c^2}\int_\infty^0 -(R^2+s^2)^{-5/2}ds.
\end{equation}
This integration yields
\begin{equation}
\label{P_mm t-coil}
\bm{P}_{mm} = -\frac{\mu_o}{4\pi}\frac{qNI\pi a^2}{R^2}\bm{\hat{k}}.
\end{equation}
In terms of the magnetic flux in the coil, this mechanical momentum is
\begin{equation}
\label{P_mm t-coil 2}
\bm{P}_{mm} = -\frac{q\Phi_B}{2R}\bm{\hat{k}}.
\end{equation}
It is seen that this is the negative of the electromagnetic momentum given in
Eq. (\ref{Pme toroid}), and once again momentum is conserved without a need
to invoke a hidden form.

\section{Discussion}

No hidden momentum is necessary to satisfy momentum conservation when a charge
is in the vicinity of either a solenoid or a toroidal solenoid. I have
calculated the momentum in these systems using the Lorentz formulation of force
and momentum in two different ways for a solenoid and for a toroidal solenoid.
First, a charge is placed in the vicinity of a coil originally containing no
magnetic flux, and then the flux is then allowed to increase to a final value.
In the second type of calculation, a charge is moved from a great distance into
the vicinity of a coil already containing flux. In neither case is hidden
momentum needed for momentum conservation. The total momentum remains zero in
all four  calculations. These calculations illustrate my assertion that the
idea of hidden momentum is incorrect and is due to erroneously overlooking what
occurs when an electric field is applied to a magnet or a magnet is created in
an electric field \cite{redfern4}. I maintain that you can't ignore the history
of charge-magnet systems and analyze them in the rest frames in which they are
found after their formation, since mechanical momentum is generated when these
systems originate.

\hfill\break

\noindent This work was done by the author without external financial support.

\hfill\break

\bibliographystyle{epj}
\bibliography{coils}

\end{document}